%% file: main.tex
\documentstyle[preprint,epsbox]{jpsj}
\newcommand{\eps}{\varepsilon}

\newcommand{\vct}[1]{\mbox{\boldmath #1}}
\def\runtitle{Magnetic Phase Transition of the Perovskite-type Ti Oxides}
\def\runauthor{Masahito {\sc Mochizuki} and Masatoshi {\sc Imada}}
\title{Magnetic Phase Transition of the Perovskite-type Ti Oxides}
\author
{Masahito {\sc Mochizuki} and Masatoshi {\sc Imada} }

\inst
{Institute for Solid State Physics, 
University of Tokyo,\\ 5-1-5 Kashiwa-no-ha, Kashiwa, Chiba 277-8581}

\recdate{\today}

\abst{
Properties and mechanism of the magnetic phase transition of the 
perovskite-type Ti oxides, which is 
driven by the Ti-O-Ti bond angle distortion, are studied theoretically
by using the effective spin and pseudo-spin Hamiltonian
with strong Coulomb repulsion.   
It is shown that the A-type antiferromagnetic(AFM(A)) to 
ferromagnetic(FM) phase transition occurs
as the Ti-O-Ti bond angle is decreased. 
Through this phase transition, the orbital state is hardly changed
so that the spin-exchange coupling along the $c$-axis 
changes nearly continuously from positive to negative 
and takes approximately zero 
at the phase boundary.
The resultant strong two-dimensionality in the spin coupling 
causes a rapid suppression 
of the critical temperature as is observed experimentally.}

\kword
{perovskite-type Ti oxides, ${\rm GdFeO}_3$-type distortion, 
$d$-level degeneracy, orbital ordering,
second-order perturbation theory, A-type antiferromagnetism, 
Mermin and Wagner's theorem}
\begin{document}
\sloppy
\maketitle
%
%
\section{Introduction} 
\begin{figure}[tdp]
  \hfil
  \epsfile{file=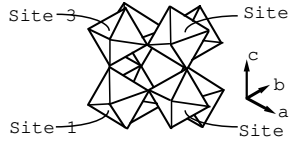,scale=0.2}
  \hfil
  \caption{${\rm GdFeO}_3$-type distortion.}
  \label{gdfo3}
\end{figure}

Perovskite-type Ti oxide $R{\rm TiO}_3$($R$ 
being a trivalent rare-earth ion)
is a typical Mott-Hubbard insulator~\cite{Imada98}.
${\rm Ti}^{3+}$ has a $3d^1$ configuration, and one of the three-fold 
$t_{2g}$ orbitals is occupied at each transition-metal 
site. The crystal structure is an orthorhombically
distorted perovskite (${\rm GdFeO}_3$-type distortion)
whose unit cell contains four octahedra
as shown in Fig.~\ref{gdfo3}.
The magnitude of the distortion depends on the ionic radii
of the $R$ ion. With a small ionic radius of $R$ ion, 
the lattice structure is more distorted and the bond angle
is decreased more largely from $180^{\circ}$.
The bond angle can 
be controlled by use of the solid-solution systems  
${\rm La}_{1-y}{\rm Y}_{y}{\rm TiO}_3$
or in $R{\rm TiO}_3$, varying the $R$ ions.
Especially, with varying the Y concentration in 
${\rm La}_{1-y}{\rm Y}_{y}{\rm TiO}_3$, we can control the 
bond angle almost continuously from $156^{\circ}$
($y = 0$) to $140^{\circ}$ ($y = 1$).
In ${\rm YTiO}_3$, a $d$-type JT distortion has been observed where the
longer and shorter Ti-O bond lengths are 
$\sim$2.08$\AA$ and $\sim$2.02$\AA$, respectively~\cite{Akimitsu98}.
Although the difference between the longer and shorter bond length
is relatively small, ${\rm LaTiO}_3$ also shows a $d$-type JT
distortion.
Recently, electronic and magnetic phase diagrams have been 
investigated intensively as functions of the magnitude of
a Ti-O-Ti bond angle distortion
~\cite{Goral82,Greedan85,Okimoto95,Katsufuji97}.
In the less distorted or a La-rich($y<0.6$)
region, the system shows AFM ground state.
In particular, ${\rm LaTiO}_3$ ($y=0.0$) shows a G-type AFM(AFM(G)) 
ground state with a magnetic moment 0.45$\mu_{\rm B}$~\cite{Goral83}.
With increasing the Y-concentration or varying the $R$ site with 
smaller size ion (decrease of the Ti-O-Ti bond angle),
the N${\rm {\grave{e}}el}$ temperature ($T_{\rm N}$)
decreases rapidly and is suppressed to almost zero,
and subsequently a FM ordering appears.
This rapid decrease of $T_{\rm N}$ is hardly explained
by the conventional models. The origin is 
one of the issues of interest.
In the relatively distorted or Y-rich region, the system shows
a FM ground state. In ${\rm YTiO}_3$($y=1.0$), the value of
the magnetic moment is 0.84$\mu_{\rm B}$~\cite{Garret81}.
This ferromagnetism is hardly explained by a simple single-band Hubbard
model and requires to consider
the $d$-level degeneracy.
Recent model Hartree-Fock studies have succeeded in reproducing the magnetic
structures of the both end compounds~\cite{Mizokawa96b,Mizokawa96a}.
The nature of the magnetic phase transition is, however, not sufficiently
clarified.
Besides, the spin and orbital states realized in the moderately
distorted region are issues of interest. 
In this paper, we study the properties and mechanism of the phase transition
by focusing on the region near the phase boundary.

\section{Formalism} 
We start with the multi-band $d$-$p$ model in which 
the full degeneracies of Ti $3d$ and O $2p$
orbitals as well as on-site Coulomb and exchange 
interactions are taken into account.
In this Hamiltonian, the effects of the ${\rm GdFeO}_3$-type distortion 
are considered through the 
$d$-$p$ transfer integrals which is defined by using the
Slater-Koster's parameters $V_{pd{\pi}}$,$V_{pd{\sigma}}$,$V_{pp{\pi}}$ 
and $V_{pp{\sigma}}$~\cite{Slater54}.
The effects of the $d$-type Jahn-Teller(JT) distortion are also
considered. 
The magnitude of the distortion is expressed
by the ratio $[V_{pd{\sigma}}^s$/$V_{pd{\sigma}}^l]^{1/3}$, 
here $V_{pd{\sigma}}^s$ and $V_{pd{\sigma}}^l$
are the transfer integrals for the shorter and longer
Ti-O bonds. 
In order to reveal the origin of the rapid suppression
of $T_{\rm N}$, we focus on the situation
near the phase boundary between AFM and FM phases.
The value of the ratio
$[V_{pd\sigma}^s/V_{pd\sigma}^l]^{1/3}$ is fixed at 
1.030, which is 
expected to be realized near the phase boundary
under the assumption of the linear increase as a function of the bond angle
from 1.036(${\rm YTiO}_3$) to 1.00(${\rm LaTiO}_3$).
Under the JT distortion, the $t_{2g}$ level-splitting energy
$\Delta_{t_{2g}}$ is estimated as 0.050eV by using Slater-Koster's relations.
Since the $t_{2g}$ level-splitting due to the spin-orbit 
interaction is sufficiently small in comparison with $\Delta_{t_{2g}}$,
we neglect the interaction 
through the present calculations. 
By integrating over the O $2p$ orbital degrees of freedom,
we derive the $effective$ Hubbard model which includes
only Ti $3d$ orbital degrees of freedom.

Under the cubic-type crystal field, a five-fold degeneracy
of $3d$ orbitals is lifted to two-fold higher levels $e_g$
and three-fold lower levels $t_{2g}$.
Moreover, in the $d$-type JT distortion, the degeneracies of the
$e_g$ and $t_{2g}$ orbitals are lifted.
Although the $e_g$ level is uniformly lifted to a higher level $3z^2-r^2$
and a lower level $x^2-y^2$
at each site, the ways of $t_{2g}$-level splitting are
different between sites 1, 3 and sites 2, 4 (see Fig. 1). 
In sites 1 and 3, the $y$-axis is elongated and consequently,
the $t_{2g}$ level is lifted to
lower $xy$ and $yz$ levels 
and a higher $zx$ level.
On the other hand, in sites 2 and 4, the $x$-axis is elongated 
so that it is lifted to 
lower $xy$ and $zx$ levels and a higher $yz$ level.
Let us represent the five $3d$ orbitals $xy$, $yz$, $zx$,
$x^2-y^2$ and $3z^2-r^2$ by energy-level indices
1, 3, 2, 4 and 5, respectively at sites 1 and 3, 
and by indices 1, 2, 3, 4 and 5
at sites 2 and 4.
As a result, in the insulating $d^1$ systems under a
JT distortion, one of the two-fold degenerate lower $t_{2g}$
orbitals is occupied at each site.
Especially, in the case of the $d$-type JT distortion,
either $xy$ or $yz$ orbital is occupied at site 1 and 3, 
$xy$ or $zx$ at site 2 and 4. 

Based on the above discussion, the multi-band Hubbard Hamiltonian
derived from the multi-band $d$-$p$ model has a form, 
\begin{equation}
        H^{\rm mH} = H_{d}^{\rm mH} + H_{tdd}^{\rm mH} 
+ H_{\rm on-site}  \\
\end{equation}
with
\begin{eqnarray}
     & &H_{d}^{\rm mH} = \sum_{i,m,\sigma} \eps_{d\,i,m}
        d_{i,m,\sigma}^{\dagger} d_{i,m,\sigma}, \\
     & &H_{tdd}^{\rm mH} = \sum_{i,m,i',m',\sigma} 
        t_{im,i'm'}^{dd}
                      d_{i,m,\sigma}^{\dagger} 
                      d_{i',m',\sigma}  + \vct{h.c.}, \\
     & &H_{\rm on-site} = H_{u} + H_{u'} + 
                            H_j + H_{j'}, 
\label{multihubhamilt}
\end{eqnarray} 
where $d_{i,m,{\sigma}}^{\dagger}$ is a creation operator of an electron 
with spin $\sigma(={\uparrow}, {\downarrow})$ in a
$3d$ level $m$ at Ti site $i$.
By $H_{d}^{\rm mH}$, we express the level energies of Ti $3d$ orbitals 
under the influence of the crystal fields with
\begin{equation}
      \eps_{d\,i,m} = \left\{
      \begin{array}{ll}
        
        \eps_{dl}                      & \mbox{for} \quad m = 1, 3, \\
        \eps_{dl} + \Delta_{t_{2g}}    & \mbox{for} \quad m = 2,  \\
        \eps_{dl} + \Delta_{x^2-y^2}   & \mbox{for} \quad m = 4,  \\
        \eps_{dl} + \Delta_{3z^2-r^2}  & \mbox{for} \quad m = 5 .\\
      \end{array} \right. 
\end{equation}
Here, $m=1,3$ are lower ${t_{2g}}$ levels,
$m=2$ is a higher ${t_{2g}}$ level and $m=4$ and $m=5$ are
lower and higher $e_g$ levels, respectively.
The $\Delta_{t_{2g}}$, $\Delta_{x^2-y^2}$ and $\Delta_{3z^2-r^2}$   
denote the level-splitting energies measured from
lower $t_{2g}$ level.
It should be noted that the same indices of energy levels 
at different sites do not necessarily correspond to the orbitals with 
the same symmetry.
$H_{tdd}^{\rm mH}$ is a $d$-$d$ super-transfer term and 
$H_{\rm on-site}$ represents on-site $d$-$d$ Coulomb interactions. 
The $H_{\rm on-site}$ term consists of the following four contributions,
\begin{eqnarray}
     & &   H_{u} = \sum_{i,m} u
        d_{i,m,\uparrow}^{\dagger} d_{i,m,\uparrow}
        d_{i,m,\downarrow}^{\dagger} d_{i,m,\downarrow}, \\
     & &   H_{u'} = \sum_{i,m>m',{\sigma},{\sigma}'} u'
        d_{i,m,\sigma}^{\dagger} d_{i,m,\sigma}
        d_{i,m',{\sigma}'}^{\dagger} d_{i,m',{\sigma}'}, \\
     & &   H_{j} = \sum_{i,m>m'\sigma,{\sigma}'} j
        d_{i,m,\sigma}^{\dagger} d_{i,m',\sigma}
        d_{i,m',{\sigma}'}^{\dagger} d_{i,m,{\sigma}'}, \\
     & &   H_{j'} = \sum_{i,m \ne m'} j'
        d_{i,m,\uparrow}^{\dagger} d_{i,m',\uparrow}
        d_{i,m,\downarrow}^{\dagger} d_{i,m',\downarrow}, 
\end{eqnarray}
where $H_{u}$ and $H_{u'}$ are the intra- and inter- orbital 
Coulomb interactions and $H_{j}$ and $H_{j'}$ denote the
exchange interactions.
The term $H_{j}$ is the origin of the Hund's rule coupling 
which strongly favors the spin alignment in the same
direction on the same atoms.
These interactions are expressed by using Kanamori parameters,
$u$, $u^{\prime}$, $j$ and $j^{\prime}$ which 
satisfy the following relations~\cite{Brandow77,Kanamori63},
$u = U + \frac{20}{9}j$, $\quad$ $u'= u -2j$
and $\quad$ $j = j'$.
Here, $U$ gives a magnitude of the multiplet-averaged 
$d$-$d$ Coulomb interaction.
The charge-transfer energy $\Delta$, which describes the energy
difference between occupied O $2p$ and unoccupied
Ti $3d$ levels, is
defined by using $U$ and energies of the
bare Ti $3d$ and O $2p$ orbitals $\eps_d^0$ and $\eps_p$
as follows,
\begin{equation}
      \Delta = \eps_{d}^0 + U -\eps_p,
\end{equation}
since the characteristic unoccupied $3d$ level energy on the singly occupied
Ti site is $\eps_{d}^0 + U$.
The values of $\Delta$, $U$ and $V_{pd\sigma}$ are
estimated by photoemission spectra~\cite{Saitoh95,Bocquet96}.
We take the values of these parameters as
$\Delta = 6.0$eV, $U = 4.0$eV, $V_{pd\sigma} = -2.0$eV
and $j = 0.74$eV
throughout the present calculation. 
The ratio $V_{pd\sigma}/V_{pd\pi}$ is fixed at $-2.17$ and
$V_{pp\sigma}$ and $V_{pp\pi}$ at 0.60eV and $-0.15$ eV, 
respectively~\cite{Harrison89}.

 Starting with the multi-band Hubbard Hamiltonian, 
we can derive an effective Hamiltonian in the low-energy region
on the subspace of states
with singly occupied $t_{2g}$ orbitals
at each transition-metal site by utilizing a 
second-order perturbation theory.
The states of $3d$ electron localized at the transition-metal
sites can be represented by two quantum numbers, the 
$z$-component of the spin $S_z$
and the number of the occupied orbitals. 
When one of the two-fold lower $t_{2g}$ orbitals
is occupied at each site, we can describe the electronic states 
using a spin-1/2 operators, which we call the pseudo-spin $\vct{$\tau$}$. 
We follow an approach similar to the well-known
Kugel-Khomskii formulation~\cite{Kugel72,Kugel73,Kugel82,Khomskii73}.
We express the $3d$ electron operators in terms of 
$\vct{$S$}$ and $\vct{$\tau$}$
to arrive at the effective spin and pseudo-spin Hamiltonian;
\begin{equation}
      H_{\rm eff} = \tilde {H}_{d}^{\rm mH} + H_{t_{2g}} + H_{e_g}.      
\end{equation}
The first term $\tilde{H}_{d}^{\rm mH}$
is obtained from the zeroth-order perturbational processes.
The second term $H_{t_{2g}}$ is
obtained from the second-order perturbational processes whose 
intermediate states contain only $t_{2g}$-orbital degrees
of freedom.
The third term $H_{e_g}$ is obtained 
from the second-order perturbational processes whose 
intermediate states contain $e_g$-orbital degrees
of freedom.

\section{Results} 
\begin{figure}[td]
  \hfil
  \epsfile{file=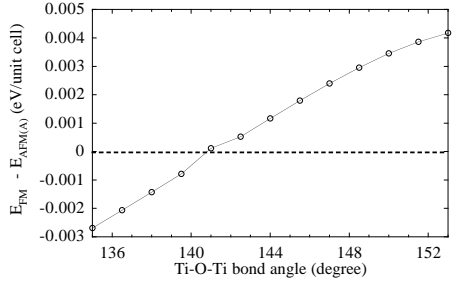,scale=0.35}
  \hfil
  \caption{Relative energy as a function of the Ti-O-Ti bond angle.}
  \label{relatene}
\end{figure}

\begin{figure}[h]
  \hfil
  \epsfile{file=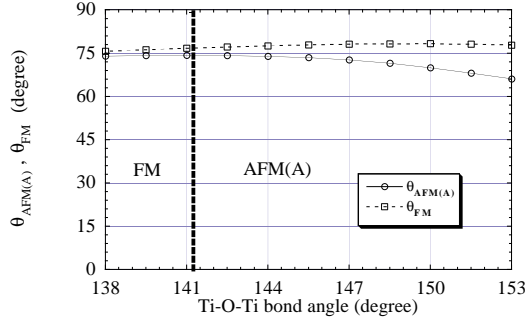,scale=0.4}
  \hfil
  \caption{Orbital states of the FM- and AFM(A) solutions.}
  \label{theta}
\end{figure}

We have calculated the total energies of various spin and orbital 
configurations by  
applying a mean-field approximation 
to the effective spin and pseudo-spin Hamiltonian.
Near the AFM-FM phase boundary, the AFM(A) and FM solutions
with a certain kind of orbital orderings are stabilized.
Figure~\ref{relatene} shows that the AFM(A) to FM phase transition 
arises by decreasing the Ti-O-Ti bond angle
(increasing the ${\rm GdFeO}_3$-type distortion).
In the AFM(A) and FM phases, $yz, zx, yz$ and $zx$ orbitals are predominantly
occupied among the two-fold degenerate lower $t_{2g}$ 
orbitals at site 1, 2, 3 and 4, respectively
(($yz,zx,yz,zx$)-type orbital order).
We can specify the orbital states realized in the AFM(A) and FM
solutions by using the angle $\theta_{\rm AFM(A)}$ and 
 $\theta_{\rm FM}$ as,
\begin{eqnarray}
&{\rm site}& 1; \quad\cos{\theta_x}|xy>+\sin{\theta_x}|yz>,
\nonumber \\
&{\rm site}& 2; \quad\cos{\theta_x}|xy>+\sin{\theta_x}|zx>,
\nonumber \\
&{\rm site}& 3; \quad -\cos{\theta_x}|xy>+\sin{\theta_x}|yz>,
\nonumber \\
&{\rm site}& 4; \quad -\cos{\theta_x}|xy>+\sin{\theta_x}|zx>,
\end{eqnarray}
where $x =$ AFM(A), FM.
In Fig.~\ref{theta}, the angles for the AFM(A) and FM solutions 
are plotted.
The difference between the $\theta_{\rm AFM(A)}$ and $\theta_{\rm FM}$
is very small, especially in the more distorted region or near
the phase boundary.
This means that the way of the orbital ordering hardly
changes through the magnetic phase transition.
Then the AFM(A) to FM phase transition is identified as
the transition where the sign of the
spin exchange interaction along the $c$-axis is changed
from negative to positive while that in the $ab$-plane
is constantly negative.
The constant FM coupling in the $ab$-plane
under the ($yz,zx,yz,zx$)-type orbital state can be easily
understood.
In the $ab$-plane, the neighboring orbitals ($yz$ and $zx$) are 
approximately orthogonal to each other.
Hence, the FM spin configuration is favored 
through Hund's coupling interaction.
However, the emergence of the FM phase is still 
controversial since the neighboring orbitals along the $c$-axis
are not orthogonal but have the same symmetries.
By considering the transfers from $yz$ to neighboring $3z^2-r^2$,
this is understood schematically as follows.

At this stage, we assume that the ($yz,zx,yz,zx$)-type
orbital order is hardly changed between two phases.
Let us consider the energy gain of an electron in the 
$yz$ orbital at site 1, which is caused by the second-order perturbational
processes with respect to the transfers along the $c$-axis 
(i.e., the transfers between site 1 and site 3). 
In the large ${\rm GdFeO}_3$-type distortion, the $yz$ orbital
at site 1 mainly hybridize with the $yz$ and $3z^2-r^2$ orbitals
at site 3 along the $z$ direction relative to the other
orbitals. 
\begin{figure}[tb]
  \hfil
  \epsfile{file=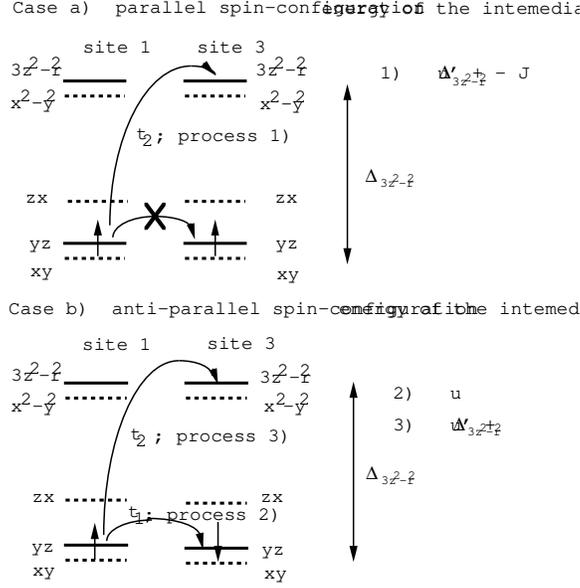,scale=0.4}
  \hfil
  \caption{Characteristic second order perturbational processes.}
  \label{sdproc}
\end{figure}
When the $yz$ orbital at site 3 is occupied by an electron,
the second-order perturbational energy gain of an electron
in the $yz$ orbital at site 1 depends on the spin configuration
between site 1 and site 3.
When the spins of electrons on site 1 and site 3 are antiparallel,
the absolute value of the energy gain can be written approximately
as follows(see Fig.~\ref{sdproc}$\quad$ a)),
\begin{equation}
       \frac{t_1^2}{u} + \frac{t_2^2}{u'+\Delta_{3z^2-r^2}}. 
\end{equation}
Here, $t_1$ represents the transfer between $yz$ at site 1
and $yz$ at site 3 and $t_2$ represents that between $yz$ at site 1
and $3z^2-r^2$ at site 3 
and $\Delta_{3z^2-r^2}$ denotes the level-energy difference
between $3z^2-r^2$ and $yz$ orbitals.
On the other hand, when the spins are parallel, transfer to the $yz$
orbital is forbidden by Pauli's principle but the 
energies of the intermediate states in which two electrons occupy 
the different orbitals are reduced by the intra-site exchange 
interaction $j$(see Fig.~\ref{sdproc}$\quad$ b)).
Consequently, the absolute value of the
energy gain can be written as 
\begin{equation}
       \frac{t_2^2}{u'+\Delta_{3z^2-r^2}-j}\sim  
  \frac{t_2^2}{u'+\Delta_{3z^2-r^2}} 
+ \frac{t_2^2 j}{(u'+\Delta_{3z^2-r^2})^2}. 
\end{equation}
Therefore, the spin configuration between site 1 and site 3 is
determined by the competition of following two energies,
$\frac{t_1^2}{u}$ and  $\frac{t_2^2 j}{(u'+\Delta_{3z^2-r^2})^2}$.
In Fig.~\ref{enegain}, the values of these energies are 
plotted as functions of the bond angle.
\begin{figure}[tb]
  \hfil
  \epsfile{file=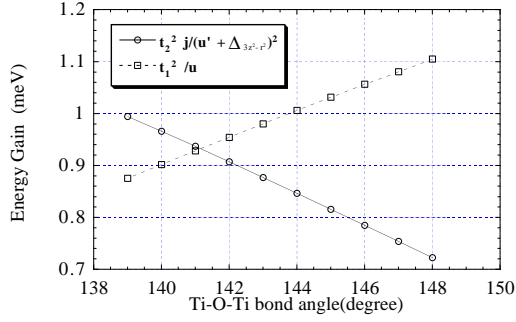,scale=0.4}
  \hfil
  \caption{Characteristic energy gains due to the second order perturbation.}
  \label{enegain}
\end{figure}
As the ${\rm GdFeO}_3$-type distortion increases, 
the indirect hybridizations between neighboring $t_{2g}$ orbitals
are decreased and those between neighboring $t_{2g}$ orbitals and
$e_g$ orbitals are increased.
Consequently, the value of $t_1^2/u$ is decreased and that of
$t_2^2 j/(u^{\prime}+\Delta_{3z^2-r^2})^2$ is increased, resulting in
crossing of two energies as the bond angle is decreased.
Moreover, since the hybridization between $t_{2g}$ and O $2p$ orbitals 
becomes to have
a $\sigma$-bonding character, the amplitudes of the $t_2$ and
$t_2^2 j/(u^{\prime}+\Delta_{3z^2-r^2})^2$ are critically
increased by the ${\rm GdFeO}_3$-type distortion.
Based on the above discussions, we can well describe this system by the
following Heisenberg model as far as the ($yz,zx,yz,zx$)-type
orbital order is strongly stabilized and hardly
affected by the change of the spin configuration,
\begin{equation}
    H_{\rm Heis} =
  J_{\rm Heis}^c \sum_{\langle i,j\rangle}^{c}    
{\vct{$S$}}_i\cdot{\vct{$S$}}_j 
 +J_{\rm Heis}^{a,b} \sum_{\langle i,j\rangle}^{a,b}
{\vct{$S$}}_i\cdot{\vct{$S$}}_j ,    
\end{equation}
with 
\begin{eqnarray}
  J_{\rm Heis}^{a,b} &<& 0 , \\
  J_{\rm Heis}^{c}
&=& 4(\frac{t_1^2}{u} - \frac{t_2^2 j}{(u'+\Delta_{3z^2-r^2})^2}).   
\end{eqnarray}
Here, $\sum_{\langle i,j\rangle}^{c}$ denotes the summation
over the neighboring spin couplings along the $c$-axis and
$\sum_{\langle i,j\rangle}^{a,b}$ in the $ab$-plane.
The AFM(A) to FM phase transition occurs by the change 
in the sign of $J_{\rm Heis}^{c}$.
Moreover, within this model, 
since the value of $J_{\rm Heis}^{c}$ decreases from
a positive value to a negative one continuously 
as the bond angle is decreased and takes zero at the phase boundary.
The two-dimensional spin coupling
is realized at the phase boundary.
Consequently, the critical temperature 
at the phase boundary suppressed to zero
according to Mermin and Wagner's theorem~\cite{Mermin66}.  
Therefore, we can conclude that a strong two-dimensionality
in spin couplings is realized near the phase boundary.

In summary, the possible scenario of the rapid decrease
of $T_{\rm N}$ is as follows.
The ${\rm GdFeO}_3$-type distortion increases the indirect hybridizations
between neighboring $t_{2g}$ and $e_g$ orbitals.
As a result, the occupancy of the orbitals directed 
along $c$-axis is strongly favored by energy gain 
due to the second-order perturbational processes with
respect to the transfers in the $z$-direction.
Hence, the ($yz,zx,yz,zx$)-type orbital order is strongly stabilized 
almost independently of the spin configuration.
Since the orbital structure is hardly changed through
the magnetic phase transition, the spin-exchange interaction
along the $c$-axis is decreased from positive value to 
negative one nearly continuously
with decreasing the bond angle, and consequently  
becomes almost zero at the phase boundary while that in the 
$ab$-plane remains constantly ferromagnetic.
This strong two-dimensionality at the phase boundary 
suppresses $T_{\rm N}$ and
$T_{\rm C}$ critically. However, a slight change of the orbital state
exist at the phase boundary, and it causes the first-order 
phase transition in this system.

\input{bib.tex}

\end{document}

%% file: bib.tex

%% file: main.bbl
\begin{thebibliography}{999}
%
%
%
\bibitem{Imada98}For a review see M. Imada, A. Fujimori and Y. Tokura:
Rev. Mod. Phys. {\bf 70} (1998) 1039.
%
\bibitem{Akimitsu98}J. Akimitsu et. al. unpublished.
%
\bibitem{Goral82}J. P. Goral, J. E. Greedan and D. A. Maclean: 
J. Solid State Chem. {\bf 43} (1982) 244.
%
\bibitem{Greedan85}J. E. Greedan, J. Less-Common Met. 
{\bf 111} (1985) 335.
%
\bibitem{Okimoto95}Y. Okimoto, T. Katsufuji, Y. Okada, T. Arima and
Y. Tokura:
Phys. Rev. B {\bf 51} (1995) 9581.
%
\bibitem{Katsufuji97}T. Katsufuji, Y. Taguchi and Y. Tokura: 
Phys. Rev. B {\bf 56} (1997) 10145.
%
\bibitem{Goral83}J. P. Goral and J. E. Greedan: 
J. Magn. Mater. {\bf 37} (1983) 315.
%
\bibitem{Garret81}J. D. Garret and J. E. Greedan:  
Inorg. Chem. {\bf 20} (1981) 1025.
%
\bibitem{Mizokawa96b}T. Mizokawa and A. Fujimori:
Phys. Rev. B {\bf 51} (1995) 12 880.
%
\bibitem{Mizokawa96a}T. Mizokawa and A. Fujimori:
Phys. Rev. B {\bf 54} (1996) 5368.
%
\bibitem{Slater54}J. C. Slater and G. F. Koster: 
Phys. Rev. {\bf 94} (1954) 1498.
%
\bibitem{Brandow77}B. H. Brandow: 
Adv. Phys. {\bf 26} (1977) 651.
%
\bibitem{Kanamori63}J. Kanamori: Prog. Theor. Phys. {\bf 30} (1963) 275.
%
\bibitem{Saitoh95}T. Saitoh, A. E. Bocquet, T. Mizokawa and A. Fujimori:
Phys. Rev. B {\bf 52} (1995) 7934.
%
\bibitem{Bocquet96}A. E. Bocquet, T. Mizokawa,
K. Morikawa, A. Fujimori, S. R. Barman, K. Mati, D. D. Sarma,
Y. Tokura and M. Onoda: Phys. Rev. B {\bf 53} (1996) 1161.
%
\bibitem{Harrison89}W. A. Harrison: Electronic Structure and the
Properties of solids (Dover, New York, 1989)
%
\bibitem{Kugel72}K. I. Kugel and D. I. Khomskii: 
Pisma. Zh. Eksp. Teor. Fiz. {\bf 15}, (1972) 629.
%
\bibitem{Kugel73}K. I. Kugel and D. I. Khomskii: Zh. Eksp. Teor. Fiz.
{\bf 64}, (1973) 1429. [Sov. Phys. JETP. {\bf 37}, (1973) 725.].
%
\bibitem{Kugel82}K. I. Kugel and D. I. Khomskii: Sov. Phys. Usp 
{\bf 25}, (1982) 231.
%
\bibitem{Khomskii73}D. I. Khomskii and K. I. Kugel: Solid State Commun.
{\bf 13}, (1973) 763.
%
%
\bibitem{Mermin66}M. D. Mermin and H. Wagner:
Phys. Rev. Lett. {\bf 17} (1966) 1133.
\end{thebibliography}
